\documentclass[prd,twocolumn]{revtex4}
\usepackage{graphicx, epsfig}
\usepackage{color}
\usepackage{mathrsfs}
\usepackage{bm}

\usepackage{graphicx,epsfig}
\usepackage{epsfig}
\usepackage{mathrsfs}
\usepackage{bm}
\usepackage{verbatim}
\usepackage{amsfonts}
\usepackage[latin1]{inputenc}
\usepackage{graphicx}
\usepackage{amssymb}
\usepackage{bm}
\usepackage{color}
\usepackage{float}
\usepackage{amsmath}
\usepackage{amsfonts}
\usepackage{dcolumn}
\usepackage{bm}
\usepackage{hyperref}
\usepackage[normalem]{ulem}
\usepackage{mathtools}

\def\be{\begin{equation}}
\def\ee{\end{equation}}
\def\ba{\begin{eqnarray}}
\def\ea{\end{eqnarray}}
\def\la{\langle}
\def\ra{\rangle}

\begin{document}
\title{Quantum Backreaction on Classical Dynamics}
\author{Tanmay Vachaspati}
\affiliation{
Physics Department, Arizona State University, Tempe, AZ 85287, USA.\\
Maryland Center for Fundamental Physics, University of Maryland, 
                    College Park, Maryland 20742, USA.
}

\begin{abstract}
\noindent
Motivated by various systems in which quantum effects occur in classical backgrounds, we consider
the dynamics of a classical particle as described by a coherent state that is coupled to a quantum bath 
via bi-quadratic interactions. We evaluate the resulting quantum dissipation of the motion of the classical 
particle. 
We also find classical initial conditions for the bath that effectively lead to the same dissipation as that 
due to quantum effects, possibly providing a way to approximately account for quantum backreaction 
within a classical analysis.
\end{abstract}

\maketitle

Several systems of interest involve the coupling of classical backgrounds to quantum fields.  
The dynamics of the classical system radiates quantum excitations and thus dissipates.
We are interested in evaluating the backreaction of the quantum excitations on the classical
dynamics.

This study is particularly relevant to gravitational systems where we do not yet have a full
quantum theory and in which context this problem has already received some attention 
\cite{Boucher:1988ua,GellMann:1992kh,Anderson:1994si,Diosi:1997mt,Yang:2017xyh}.
For example, in inflationary cosmology, classical 
dynamics of the inflaton field excites quantum fields that then become observable 
cosmological density perturbations.
The inflaton field denoted $\Phi(t)$ is assumed to be homogeneous and initially displaced from
its minimum. As the field rolls towards its minimum, it can excite a second field, $\phi$, that is
coupled to it. Generally symmetries under $\Phi \to -\Phi$ and $\phi \to -\phi$ are assumed
so that the lowest order coupling term is $\lambda \Phi^2 \phi^2$.
The classical evolution of $\phi$ will be governed by
\be
\square \phi + m^2 \phi + 2\lambda \Phi^2 \phi =0
\label{phieq}
\ee
and the initial condition $\phi=0$, ${\dot \phi}=0$, gives
$\phi=0$ for all times. In quantum theory, however, if $\phi$ is assumed in
its ground state initially, it gets excited by the dynamics of the $\Phi$ field. Then the 
quantum evolution
of $\phi$ is non-trivial and it backreacts on the dynamics of $\Phi$ and dissipates its
motion. We are interested in evaluating this quantum dissipation. We are also interested
in finding a set of classical initial conditions different from $\phi=0={\dot\phi}$ for which the 
classical dissipation closely agrees with the quantum result.

These questions are of interest beyond inflationary cosmology. 
Gravitational collapse leads to Hawking radiation that is purely quantum and 
this will cause the collapsing body to evaporate. The collapsing body is a large object that is 
most conveniently treated clasically, as is its gravitational field. But the radiation is 
quantum. Can the backreaction on the collapse be estimated on the basis of a classical 
calculation? 

There are non-gravitational settings where similar questions arise. For example,
what is the backreaction of Schwinger pair production on the electric field? A full treatment
of this problem in 1+1 dimensions for the special case of massless fermions leads
to an interesting $t^{-1/2}$ decay of the electric field and an effective electrical
conductivity of the vacuum \cite{Chu:2010xc} but the case of massive fermions is
still open. Another setting where classical and quantum descriptions confront each other is
when discussing the production of topological solitons in particle 
collisions \cite{Dutta:2008jt,Vachaspati:2016abz}.
Solitons are solutions of the classical field theory equations and this is the most convenient
framework to discuss them. In studying the creation of solitons by scattering particles, if the 
initial condition involves a large number of particles, they too can be described by classical 
equations. Thus one
may be inclined to think that classical evolution is sufficient to study the creation of
solitons in (many) particle collisions. However this is not true in general because,
depending on the initial conditions, the classical evolution may be restricted to an 
embedded subspace of the model \cite{Vachaspati:1992pi,Barriola:1993fy}, just as $\phi=0$ 
is the dynamical subspace in the example of Eq.~(\ref{phieq}). Solitons, by their topological nature, 
involve a very large part of the dynamical space of field configurations and, in certain situations,
quantum effects could be crucial for the dynamics to explore the full space of fields
necessary to create solitons.

A concrete example helps to explain this issue better. Consider light on light
collisions. These involve the collisions of a large number of photons and a classical
description via Maxwell's equations should suffice. However, then the collision
is trivial since Maxwell's equations are linear. In quantum theory, 
photon collisions will sometimes produce charged particle-antiparticle
pairs ({\it e.g.} $W^\pm$, electrons, and other standard model particles).
These will create a plasma that will backreact on the dynamics of the
light on light collisions. Only the quantum dynamics will explore the full standard 
model and possibly produce electroweak strings \cite{Vachaspati:1992fi} or 
sphalerons \cite{Manton:1983nd} that are solutions of the classical electroweak equations.

The problem outlined above is very difficult to address in field theory and we will
only solve a simpler quantum mechanical problem. We first expand the fields in modes. 
For example for a scalar field,
\be
\phi(t,{\bf x}) = \sum_{\bf k} c_{\bf k}(t) f_{\bf k}({\bf x})
\ee
where $f_{\bf k}({\bf x})$ are a set of orthonormal mode functions, $c_{\bf k}(t)$ 
are mode coefficients, and the sum is an integral if the modes form a continuum. 
Then, as is standard in quantum field theory (for example see \cite{Peskin:1995ev}), 
the free field part of the theory is equivalent to an infinite set of simple harmonic 
oscillators (SHOs) given by the variables $c_{\bf k}(t)$ and these can be quantized. 
The interaction terms in the field theory lead to couplings between the modes and 
are equivalent to couplings between the SHOs. An interaction term of the type 
$\lambda \Phi^2\phi^2$, as discussed above, will be equivalent to coupling four 
SHOs, two corresponding to mode coefficients of $\Phi$ and two to those of $\phi$.
In general the couplings will be of the form $C_{{\bf k}_1} C_{{\bf k}_2} c_{{\bf k}_3} c_{{\bf k}_4}$ 
with ${\bf k}_1+{\bf k}_2+{\bf k}_3+{\bf k}_4=0$, where $C_{\bf k}$ denotes a mode 
coefficient of $\Phi$. The bi-quadratic terms, $C_{\bf K}^2 c_{\bf k}^2$, are the only 
ones that are symmetric under $C_{{\bf k}_1} \to - C_{{\bf k}_1}$ and also, separately,
$c_{{\bf k}_3} \to - c_{{\bf k}_3}$ and hence are the only ones that will survive if we evaluate 
the expectation value of the coupling term. This suggests that the bi-quadratic couplings may 
dominate and our simplification in what follows will be to only consider this coupling. However, 
this simplification should be examined further because there are many more terms that are 
not bi-quadratic and fluctuations, not just the expectation value, may be important.
(Systems with bi-linear couplings, $C_{\bf K} c_{\bf k}$, can be diagonalized 
and have been analyzed in early work~\cite{Feynman:1963fq,Caldeira:1981rx}.) 
Since $C_{\bf K}$ represents a classical degree of freedom, we take it to be in a coherent 
state initially in our quantum analysis, while $c_{\bf k}$'s are quantum variables that are 
taken to be in their ground state initially. 

To summarize this discussion, we consider a heavy SHO
coupled to a bath of light SHOs via bi-quadratic couplings. 
A solution of the classical equations is that the heavy SHO oscillates and the light SHOs 
remain at rest. This picture changes in the quantum analysis in which the heavy SHO is 
initially described by a coherent state and the light SHOs are in their ground state.
Oscillations of the heavy SHO excite the light SHOs and there are two forms of backreaction
on the heavy SHO. First the heavy SHO motion gets damped. Second, the state of the
heavy SHO is no longer a coherent state and the heavy SHO state changes towards becoming
less classical, more quantum. In the present paper we focus on the backreaction that
causes dissipation. The backreaction that takes the heavy SHO out of its coherent
state is interesting but not directly relevant to the dynamical question and we postpone
it for the time being. 

We start out by describing the quantum mechanical model in Sec.~\ref{model}.
Sec.~\ref{classical} contains our classical analysis which we perform with action-angle
variables, first studying the dynamics 
for a single light SHO, followed by a calculation of the classical dissipation for a bath of 
SHOs. The bath is essential to obtain dissipation because otherwise there is energy
exchange between the heavy and light SHOs but no dissipation. In Sec.~\ref{quantum} 
we analyze the quantum model, first for a single light SHO, then for a bath of light SHOs, 
and we then evaluate the quantum dissipation.
Our final result for the quantum vs. classical backreaction is discussed in 
Sec.~\ref{comparison} and the reader who is not interested in the details
of the calculations can directly go to Sec.~\ref{comparison}. We conclude in 
Sec.~\ref{conclusions}. Appendix~\ref{quantumphasevariables} contains a discussion 
of quantization of the SHO using action-angle variables.

%
%

\section{Model}
\label{model}

The heavy SHO position and momentum variables are $(X,P)$; the light SHO variables
are $(x_i,p_i)$ for $i=1,\ldots, N$. Traditionally, we would write the Hamiltonian
\ba
H &=& \frac{P^2}{2M}+\frac{1}{2}M \Omega^2 X^2 +
                 \sum_{i=1}^N\left ( \frac{p_i^2}{2m_i}+\frac{1}{2} m_i \omega_i^2 x_i^2 \right )
                 \nonumber \\ && \hskip 2 cm
+ \frac{1}{2N} X^2 \sum_{i=1}^N \frac{\epsilon_i}{l_i^4} x_i^2
\ea
where $l_i$ is a length scale and $\epsilon_i$ has dimensions of energy.
Rescaling
\ba
(M\Omega)^{1/2} X \to X, \ \ (m_i\omega_i)^{1/2} x_i \to x_i, \\
P \to (M\Omega)^{1/2} P, \ \ p_i \to (m_i\omega_i)^{1/2} p_i 
\ea
and assuming a universal coupling, {\it i.e.} $\epsilon_i /l_i^4$ are independent of $i$,
and dividing throughout by a factor of $\Omega$,
we get the Hamiltonian in the form
\be
H = \frac{P^2}{2}+\frac{X^2}{2}+\sum_{i=1}^N \omega_i \left ( \frac{p_i^2}{2}+\frac{x_i^2}{2} \right )
+ \frac{\epsilon}{2N} X^2 \sum_{i=1}^N x_i^2
\ee

Note: We do {\it not} use the Einstein summation convention.

\section{Classical analysis}
\label{classical}

\subsection{Single light SHO}
\label{singlelightsho}

A neat method to do the classical calculation is to perform a canonical transformation so
that the phase of the SHO is the coordinate variable and the amplitude is related to the 
momentum variable,
\be
q \to \sqrt{\frac{2I}{m\omega}} \sin\theta, \ \ 
p \to \sqrt{2Im\omega} \cos\theta .
\ee
The new Hamiltonian is
\be
H_{\rm new} = I_1 + \omega I_2 + 2 \epsilon I_1 I_2 \sin^2\theta_1 \sin^2\theta_2
\ee
where $(\theta_1,I_1)$ are variables for the heavy SHO and $(\theta_2,I_2)$
are for the light SHO. The equations of motion are
\ba
{\dot \theta}_1 &=& 1 + 2 \epsilon I_2 \sin^2\theta_1 \sin^2\theta_2 
\nonumber \\
{\dot I}_1 &=& - 2 \epsilon I_1 I_2 \sin(2\theta_1) \sin^2\theta_2 
\nonumber \\
{\dot \theta}_2 &=& \omega + 2 \epsilon I_1 \sin^2\theta_1 \sin^2\theta_2 
\nonumber \\
{\dot I}_2 &=& - 2 \epsilon I_1 I_2 \sin^2\theta_1 \sin(2\theta_2) 
\ea
The unperturbed solution (with $\epsilon \to 0$) is
\ba
\theta_1 &=& t + \phi_1
\nonumber \\
I_1 &=& K_1
\nonumber \\
\theta_2 &=& \omega t + \phi_2 
\nonumber \\
I_2 &=& K_2
\ea
where $\phi_1$, $\phi_2$, $K_1$ and $K_2$ are constants.

To first order in $\epsilon$,
\ba
\theta_1 &=& 
 t + \phi_1 \nonumber \\ && \hskip -0.7 cm
 + 2\epsilon K_2 \int_0^t dt' \, \sin^2( t' + \phi_1) \sin^2(\omega t'+\phi_2) 
 \label{firstordersolntheta1} \\
I_1 &=& K_1 
 \nonumber \\ && \hskip -0.7 cm
- 2\epsilon K_1 K_2 \int_0^t dt' \, \sin(2(t'+\phi_1)) \sin^2(\omega t'+\phi_2)
\label{firstordersolnI1} \\
\theta_2 &=&
\omega t + \phi_2 
 \nonumber \\ && \hskip -0.7 cm
+ 2\epsilon K_1 \int_0^t dt' \,  \sin^2(t'+\phi_1) \sin^2(\omega t'+\phi_2) 
\label{firstordersolntheta2} \\
I_2 &=& K_2 
 \nonumber \\ && \hskip -0.7 cm
- 2\epsilon K_1 K_2 \int_0^t dt' \, \sin^2(t'+\phi_1) \sin(2(\omega t' +\phi_2)) 
\label{firstordersolnI2}
\ea

To connect with the the usual position of the heavy SHO we use
\ba
X &&= \sqrt{2I_1}\sin\theta_1 
\nonumber \\ && \hskip -1 cm
= \sqrt{2K_1}
 \biggl [ 1 - 2\epsilon K_2 \int_0^t dt' \, \sin(2(t'+\phi_1)) \sin^2(\omega t'+\phi_2)  \biggr ]^{1/2}
\nonumber \\ &&\hskip -1 cm
\times \sin\biggl [  t + \phi_1 + 2\epsilon K_2 \int_0^t dt' \, \sin^2( t' + \phi_1) \sin^2(\omega t'+\phi_2) \biggr ]
\ea
In terms of the oscillation amplitudes, $X_0$ and $A$, we take $K_1= X_0^2/2$,
$K_2 =A^2/2$.
If the initial condition is that the heavy SHO is displaced but at rest, we take
$\phi_1=\pi/2$; for the phase of the light SHO we write $\phi_2=\phi$. Then,
\ba
X &=& X_0 \biggl [ 1 + \epsilon A^2 \int_0^t dt' \, \sin(2t') \sin^2(\omega t'+\phi)  \biggr ]^{1/2}
\nonumber \\ && \hskip -0.5 cm
\times \cos\biggl [  t + \epsilon A^2 \int_0^t dt' \, \cos^2( t') \sin^2(\omega t'+\phi) \biggr ]
\label{canonicalX}
\ea
These integrals can be done in closed form but the expressions are not illuminating.


The modified frequency of oscillation can be found by identifying the linearly growing phase
of the cosine in Eq.~(\ref{canonicalX}) and is obtained by using
\be
\int_0^t dt' \, \cos^2( t') \sin^2(\omega t'+\phi) = \frac{t}{4} + \, {\rm oscillating\ terms}.
\ee
This gives the oscillation frequency to first order in $\epsilon$,
\be
\Omega = 1 + \frac{\epsilon}{4} A^2.
\label{OmegaclassicalN=1}
\ee
In Sec.~\ref{comparison} we will find $A$ for which this modified frequency agrees with the modified 
frequency in the quantum analysis. 

\subsection{Classical dissipation for bath of light SHOs}
\label{classdiss}

To obtain dissipation we have to work out ${\dot I}_i$ to second order in $\epsilon$.
In the equation,
\ba
{\dot I}_1 &=& - 2 \epsilon I_1 I_2 \sin(2\theta_1) \sin^2\theta_2 
\ea
we insert the first order expressions in Eq.~(\ref{firstordersolntheta1})-(\ref{firstordersolnI2}). It is convenient to
define
\ba
J &\equiv& - \frac{t}{4} + \int_0^t dt' \sin^2 (t'+\phi_1) \sin^2(\omega t' + \phi_2)
\nonumber \\ 
&=& - \frac{[ \sin(2\alpha)-\sin(2\phi_1)]}{8} - \frac{[\sin(2\beta)-\sin(2\phi_2)]}{8\omega}
\nonumber \\ && \hskip 0.5 cm
+\frac{[\sin(2(\alpha+\beta))-\sin(2\phi_+)]}{16(1+\omega)}
\nonumber \\ && \hskip 0.5 cm
+\frac{[\sin(2(\alpha-\beta))-\sin(2\phi_-)]}{16(1-\omega)}
\ea
where $\alpha=t+\phi_1$, $\beta=\omega t+\phi_2$, and $\phi_\pm=\phi_1\pm \phi_2$.

Then
\be
\frac{\partial J}{\partial\phi_1} = \int_0^t dt' \sin (2(t'+\phi_1)) \sin^2(\omega t' + \phi_2)
\ee
\be
\frac{\partial J}{\partial\phi_2} = \int_0^t dt' \sin^2 (t'+\phi_1) \sin (2(\omega t' + \phi_2))
\ee
and
\ba
{\dot I}_1 
&=& -2\epsilon K_1 K_2
\left [ 1- 2\epsilon \left ( K_1 \frac{\partial J}{\partial\phi_2} + K_2 \frac{\partial J}{\partial\phi_1} \right ) \right ]
\nonumber \\ && 
\times [ \sin (2\alpha' ) + 4\epsilon K_2 J \cos (2\alpha ') ]
\nonumber \\ && 
\times [ \sin^2 (\beta ')+2\epsilon K_1 J \sin (2\beta ') ]
\label{dotI}
\ea
where $\alpha' =  (1+\epsilon K_2/2)t+\phi_1$, $\beta' = (\omega +\epsilon K_1/2)t+\phi_2$.
We have discarded terms of higher order than $\epsilon^2$ except to show linear order 
corrections to the oscillation frequencies even if these corrections lead to higher order corrections
in ${\dot I}_1$.

We ignore the order $\epsilon$ terms since they are oscillating and do not lead to dissipation.
With some algebra
\ba
{\dot I}_1 &\to& \epsilon^2 4 K_1K_2 \biggl [ 
\frac{1}{2}\left ( K_1 \frac{\partial J}{\partial \phi_2} + K_2 \frac{\partial J}{\partial \phi_1} \right )
\nonumber \\ && \hskip 1 cm
\times \left \{ \sin (2\alpha )- \frac{1}{2}\sin(2\alpha_+) - \frac{1}{2}\sin(2\alpha_-)\right \}
\nonumber \\ && \hskip -1.2 cm
+ J \left \{ \frac{K_+}{2} \cos (2\alpha_+) - \frac{K_-}{2} \cos (2\alpha_-) - K_2 \cos (2\alpha) \right \} \biggr ]
\ea
where $K_\pm = K_1\pm K_2$ and $\alpha_\pm = \alpha\pm\beta = (1\pm \omega)t+\phi_\pm$
with $\phi_\pm =\phi_1\pm \phi_2$.

We want to find the dissipation when the classical SHO is coupled to a bath of independent, incoherent,
light SHO's.
Let us assume that the bath of light SHO's has a spectral distribution of frequencies given by
a function $n(\omega)$. In other words, the number of light SHO's with frequencies between
$\omega$ and $\omega+d\omega$ is $n(\omega) d\omega$. Therefore we will calculate
\be
{\dot E}_{1,{\rm classical}} \equiv \la {\dot I} \ra = \int_0^\infty d\omega ~ n(\omega) {\dot I}
\label{dotE1defn}
\ee
Further, we are only interested in the dissipatory terms, not in the oscillatory terms. We will 
also assume $n(0)=0$. Then the terms that dominate have $(1-\omega)$ in the denominator 
and we can effectively replace
\ba
J &&\to \frac{\sin(2((1-\omega)t+\phi_-))-\sin(2\phi_-)}{16(1-\omega)}
\nonumber \\ && \hskip -0.3 cm
= - \frac{\sin^2 ((1-\omega)t) \sin(2\phi_-)}{8(1-\omega)} + \frac{\sin (2(1-\omega)t)\cos(2\phi_-)}{16(1-\omega)}
\nonumber \\ && \hskip -0.3 cm
\to  \frac{\sin (2(1-\omega)t)\cos(2\phi_-)}{16(1-\omega)}
\label{Jfinal}
\ea
since, in the last step, the first term tends to zero as $1-\omega \to 0$, while
the second term goes to a finite value. Similarly
\be
\frac{\partial J}{\partial\phi_1} 
\to - \frac{\sin (2(1-\omega)t)\sin(2\phi_-)}{8(1-\omega)} = -2J \tan (2\phi_-)
\ee
\be
\frac{\partial J}{\partial\phi_2} \to + \frac{\sin (2(1-\omega)t)\sin(2\phi_-)}{8(1-\omega)} 
= +2J \tan(2\phi_-)
\ee
Recognizing that the integration over $\omega$ in Eq.~(\ref{dotE1defn}) will be dominated by
$\omega \approx 1$ and that the oscillating terms do not contribute to the dissipation, we obtain
\be
{\dot I}_1 \to \frac{\epsilon^2}{8} K_1 K_2 K_-  \frac{\sin (2(1-\omega)t)}{(1-\omega)}
\ee
where we have replaced $J$ using Eq.~(\ref{Jfinal}).
Next we use
\be
\int_0^\infty dx \, \frac{\sin(x-x_0)}{x-x_0} \approx  
\int_{-\infty}^\infty dx \, \frac{\sin(x-x_0)}{x-x_0} = \pi
\ee
for $x_0 \gg 1$, and get
\be
{\dot E}_{1,{\rm classical}} \approx 
- \epsilon^2 \frac{\pi}{8} K_1 K_2 K_-  n(1)
\ee
for $t \gg 1$.
In terms of the initial amplitudes of the SHO's, we take $K_1= X_0^2/2$, $K_2=A^2/2$,
to get
\be
{\dot E}_{1,{\rm classical}} \approx 
-\frac{\pi}{64} \epsilon^2 n(1) X_0^4 A^2 \left (1 - \frac{A^2}{X_0^2} \right )
\label{dotE1classical}
\ee
where $A$ is the amplitude of the bath of SHO's at the resonant frequency $\omega=1$.
A surprising feature of this result is that the phases of the SHOs have dropped out.

\section{Quantum analysis}
\label{quantum}

The action-angle variables $(\theta, I)$ used in the classical analysis were more
convenient as they enabled a direct calculation of the change in the energy of the
heavy SHO due to backreaction. Quantization in these variables
is described in Appendix~\ref{quantumphasevariables} and is subtle because of
operator ordering issues. Also, since the perturbation term involves the SHO
positions, action-angle variables do not lead to any obvious simplifications in the
quantum analysis and we 
work with the conventional $(x,p)$ coordinates.

Write the wavefunction in SHO Fock basis states
\be
\psi (t, X, x)=\sum_{n,m=0}^\infty c_{nm}(t) f_n(t) \, |n\ra_X |m\ra_x
\ee
where
\be
f_n(t) = e^{-it/2} e^{-|z|^2/2} \frac{z^n}{\sqrt{n!}} 
         = e^{-it/2} e^{-|z_0|^2/2} \frac{z_0^n e^{-int}}{\sqrt{n!}}  .
\ee
In the second equality, we have used the coherent state solution $z = z_0 e^{-it}$.

The initial state is taken to be a direct product of a coherent state for $X$ and 
ground state for $x$, {\it i.e.},
\be
c_{nm}(0) = \delta_{m0}.
\label{initialcondition}
\ee
For convenience, we shall also use the notation
\be
b_{nm}(t)=c_{nm}(t) f_n(t).
\ee

In terms of creation and annihilation operators
\be
A = \frac{1}{\sqrt{2}}(X+iP), \ \ A^\dag = \frac{1}{\sqrt{2}}(X-iP), 
\label{Adefn}
\ee
\be
a = \frac{1}{\sqrt{2}}(x+ip), \ \ a^\dag = \frac{1}{\sqrt{2}}(x-ip)
\label{adefn}
\ee
we have
\ba
H = && \left ( A^\dag A+\frac{1}{2} \right ) 
+\omega \left ( a^\dag a+\frac{1}{2} \right ) 
\nonumber \\ && \hskip 1 cm
+ \frac{\epsilon}{2} \left ( \frac{A^\dag+A}{\sqrt{2}} \right )^2 
\left ( \frac{a^\dag+a}{\sqrt{2}} \right )^2
\ea
Then the Schrodinger equation gives
\ba
i \partial_t b_{nm} &=& 
\left [ \left ( n + \frac{1}{2} \right ) + \omega \left ( m + \frac{1}{2} \right ) \right ] b_{nm}
\nonumber \\ && \hskip -1 cm
+ \frac{\epsilon}{8} \sum_{l,k=0}^\infty 
\la n | ( A^\dag+A )^2 | l \ra \la m | (a^\dag+a )^2 | k \ra \, b_{lk}
\ea
Now use
\ba
\la n | ( A^\dag+A )^2 | l \ra &=& \sqrt{n(n-1)} \delta_{n,l+2} + (2n+1) \delta_{n,l} 
\nonumber \\ && \hskip 0.5 cm
+\sqrt{(n+2)(n+1)}\delta_{n,l-2}
\label{nX2l}
\ea
\ba
\la m | (a^\dag+a )^2 | k \ra &=& \sqrt{m(m-1)} \delta_{m,k+2} + (2m+1) \delta_{m,k} 
\nonumber \\ && \hskip 0.5 cm
+\sqrt{(m+2)(m+1)}\delta_{m,k-2}
 \label{nx2l}
 \ea
to get
\ba
i \partial_t b_{nm} &=& 
\biggl [ \left ( n + \frac{1}{2} \right ) + \omega \left ( m + \frac{1}{2} \right ) 
\nonumber \\ && \hskip 2 cm
+\frac{\epsilon}{8} (2n+1)(2m+1) \biggr ] b_{nm}
\nonumber \\ && \hskip -2 cm
+ \frac{\epsilon}{8} \biggl [
\sqrt{n(n-1)} \, \{ \sqrt{m(m-1)} b_{n-2,m-2} 
\nonumber \\ && \hskip -1 cm
+ (2m+1) b_{n-2,m} + \sqrt{(m+2)(m+1)} b_{n-2,m+2} \}
\nonumber \\ && \hskip -1.5 cm
+ (2n+1) \{  \sqrt{m(m-1)} b_{n,m-2} 
\nonumber \\ && \hskip 1.5 cm
+ \sqrt{(m+2)(m+1)} b_{n,m+2} \}
\nonumber \\ && \hskip -1.5 cm
+\sqrt{(n+2)(n+1)} \{ \sqrt{m(m-1)} b_{n+2,m-2} 
\nonumber \\ && \hskip -1.5 cm
+ (2m+1) b_{n+2,m} + \sqrt{(m+2)(m+1)} b_{n+2,m+2} \}
\biggr ]
\ea
Note that this equation for $b_{nm}$ also has a term proportional to $b_{nm}$ on
the right-hand side. This term is responsible for changing the frequency of 
oscillations and is better brought over to the left-hand side leading to,
\ba
\partial_t \left ( e^{i {\tilde E}_{nm} t} b_{nm} \right ) &=& 
\nonumber \\ && \hskip -3 cm
- i \frac{\epsilon}{8} e^{i {\tilde E}_{nm} t} 
\biggl [ \sqrt{n(n-1)} \biggl \{ \sqrt{m(m-1)} b_{n-2,m-2} 
\nonumber \\ && \hskip -2.5 cm
+ (2m+1) b_{n-2,m} 
+ \sqrt{(m+2)(m+1)} b_{n-2,m+2} \biggr \}
\nonumber \\ && \hskip -2 cm
+ (2n+1)\biggl  \{  \sqrt{m(m-1)} b_{n,m-2} 
\nonumber \\ && \hskip 0 cm
+ \sqrt{(m+2)(m+1)} b_{n,m+2} \biggr \}
\nonumber \\ && \hskip -3 cm
+\sqrt{(n+2)(n+1)} \biggl \{ \sqrt{m(m-1)} b_{n+2,m-2} 
\nonumber \\ && \hskip -3 cm
+ (2m+1) b_{n+2,m} + \sqrt{(m+2)(m+1)} b_{n+2,m+2} \biggr \}
\biggr ]
\label{bnmmaster}
\ea
where
\be
{\tilde E}_{nm} \equiv \left ( n + \frac{1}{2} \right ) + \omega \left ( m + \frac{1}{2} \right ) 
+ \frac{\epsilon}{8} (2n+1)(2m+1)
\ee

Eq.~(\ref{bnmmaster}) is our master equation for $b_{nm}(t)$ that we will solve perturbatively.

\subsection{Perturbative treatment of single light SHO case}
\label{pertN=1}

To first order in $\epsilon$, we can replace $b_{lk}$ on the right-hand side of Eq.~(\ref{bnmmaster})
by its unperturbed value
\be
b_{nm} = f_n(t) e^{-i\omega t/2} \delta_{m0} + {\cal O}(\epsilon )
\ee
 to get
 \ba
\partial_t \left ( e^{i {\tilde E}_{nm} t} b_{nm} \right ) &=& 
\nonumber \\ && \hskip -3 cm
- i \frac{\epsilon}{8} e^{i ({\tilde E}_{nm}-\omega/2) t} 
\biggl [ \left ( z^2 + (2n+1)+\frac{n(n-1)}{z^2} \right ) \sqrt{2} \delta_{m,2} 
\nonumber \\ && \hskip 0 cm
+ \left ( z^2 +\frac{n(n-1)}{z^2} \right ) \delta_{m,0}
\biggr ] f_n 
\label{bnmmasterpert}
\ea


Therefore only $b_{n0}$ and $b_{n2}$ are non-trivial. For $b_{n0}$ we get
\ba
b_{n0}(t) &=& e^{-i\omega t/2} \biggl [ e^{-i\epsilon (2n+1)t/8} 
\nonumber \\ && \hskip -1.2 cm
- i \frac{\epsilon}{8} 
\left \{ z_0^2 e^{-it} + \frac{n(n-1)}{z_0^2} e^{+it} \right \} \sin (t)
\biggr ] f_n(t)
\label{bn0master}
\ea
Note that a perturbation expansion in powers of $\epsilon$ would mean that we series 
expand the $\exp(-i\epsilon (2n+1)t/8)$ term. However, then there is a term that is
linear in $t$ and the expansion is valid only for very short times, in fact in an $n$
dependent way. The way we have done 
the calculation here separates out changes in the frequency of oscillation and then the 
result is valid for all times, as we have also seen in the classical case. Also, we will
see that although the correction in Eq.~(\ref{bn0master}) has a term that goes
like $\epsilon n (n-1)/z_0^2$, this contribution is of the same order (and cancels) the
term that goes like $z_0^2$. 

Another peculiarity is that the correction term to $b_{n0}$ does not vanish when
$z_0=0$ if $n=2$. This suggests that even if the heavy SHO coherent state is not 
oscillating, it will excite the second SHO. This can be seen directly from Eq.~(\ref{bnmmaster})
in which the term $(2m+1)b_{n-2,m}$ is non-zero for $n=2$, $m=0$ even if 
$z_0=0$ because $f_{n-2}=1$ for $n=2$ and $z_0=0$.  Excitations of the
light SHO in the background of a static coherent state are to be expected since
the chosen initial state is an eigenstate only of the unperturbed Hamiltonian,
not of the full Hamiltonian.

The solution for $b_{n2}$ is
\ba
b_{n2}(t) &=& 
-i \frac{\epsilon}{4\sqrt{2}}e^{-i3\omega t/2} 
\nonumber \\ && \hskip -1.5 cm
\times \biggl [   e^{-it}  z_0^2 \frac{\sin((\omega-1)t)}{\omega-1} + (2n+1)\frac{\sin(\omega t)}{\omega}
\nonumber \\ && \hskip 0 cm
+ e^{+it}  \frac{n(n-1)}{z_0^2} \frac{\sin ((\omega+1)t)}{\omega+1} \biggr ] f_n (t)
\label{bn2master}
\ea

\subsection{Expectation values}
\label{expectationvaluesN=1}

\subsubsection{Energy of heavy SHO}
\label{H1}

The Hamiltonian of the heavy SHO is
\be
H_1 = A^\dag A + \frac{1}{2}
\ee
We will calculate the time derivative of $\la H_1 \ra$,
\be
\frac{d\ }{dt} \la H_1 \ra = i \la [ H, H_1] \ra
\ee
Now
\be
[ H, H_1] = \frac{\epsilon}{2} x^2 [ X^2, A^\dag A] 
= \frac{\epsilon}{2} x^2 ( A^2 - (A^\dag)^2 )
\ee
We use
\ba
\la n | A^2-(A^\dag)^2 | l \ra = && 
\nonumber \\ && \hskip -2.5 cm
\sqrt{(n+1)(n+2)} \delta_{n+2,l} - \sqrt{n(n-1)} \delta_{n-2,l}
\ea
\be
\la 0 |x^2 |0\ra =\frac{1}{2}, \ \
\la 0 |x^2 |2\ra =\frac{1}{\sqrt{2}} = \la 2 |x^2 |0\ra
\ee
Therefore
\ba
\frac{d\ }{dt} \la H_1 \ra &=& - \frac{\epsilon}{2} \sum_{n} \sqrt{(n+1)(n+2)}
\nonumber \\ && \hskip -2 cm
\times {\rm Im} \biggl [ b^*_{n,0} b_{n+2,0} 
+ \sqrt{2} \biggl ( b^*_{n,0} b_{n+2,2} + b^*_{n,2} b_{n+2,0} \biggr ) \biggr ] 
\label{dH1dt}
\ea
We need the coefficients $b_{n,m}$ only to first order in $\epsilon$ to get the time
derivative of $\la H_1 \ra$ to second order in $\epsilon$.

Insert $b_{n,0}$ and $b_{n,2}$ from Eqs.~(\ref{bn0master}) and (\ref{bn2master})
to obtain
\ba
\sum_n \sqrt{(n+1)(n+2)} \, {\rm Im}(b^*_{n0} b_{n+2,0})  &=& 
\nonumber \\ && \hskip -5 cm
-z_0^2 \sin \left ( (2+\epsilon/2) t \right ) - \frac{\epsilon}{8}(2z_0^2+1) \sin (2t)  
\label{firstterm}
\ea
where we have used
\be
\sum_n |f_n|^2=1, \ \ \sum_n n |f_n|^2 = z_0^2, \ \ \sum_n n(n-1) |f_n|^2 = z_0^4,
\label{sumovern}
\ee
that can be derived from the identity,
\be
\left ( x \frac{d}{dx} \right )^k e^x = \sum_{n=0}^\infty n^k \frac{x^n}{n!}.
\ee

Next we calculate the middle term on the right-hand side of Eq.~(\ref{dH1dt}) 
\ba
\sqrt{2} \sum_n \sqrt{(n+1)(n+2)} \, {\rm Im}(b^*_{n,0} b_{n+2,2})  &=& 
\nonumber \\ && \hskip -7 cm
-\frac{\epsilon}{8} \biggl [ 
\frac{ \{ (4z_0^2+1)(z_0^2+2)\omega^2-2(2z_0^2+1)\omega-(2z_0^2+5)z_0^2 \} }{\omega(\omega^2-1)}
\nonumber \\ && \hskip -3 cm
\times \sin(2(\omega+1)t)
\nonumber \\ && \hskip -5 cm
+ \frac{(2z_0^2+5)z_0^2}{\omega}\sin(2t)+\frac{z_0^4}{\omega-1}\sin(4t) \biggr ]
\ea
and the final term of Eq.~(\ref{dH1dt}) is,
\ba
\sqrt{2} \sum_n \sqrt{(n+1)(n+2)} \, {\rm Im}(b^*_{n,2} b_{n+2,0})  &=& 
\nonumber \\ && \hskip -6 cm
\frac{\epsilon}{8} z_0^2 \biggl [ 
\frac{ \{ (4z_0^2+1) \omega^2-(2z_0^2+1)\} }{\omega (\omega^2-1)} \sin(2(\omega-1)t)
\nonumber \\ && \hskip -5 cm
- \frac{(2z_0^2+1)}{\omega}\sin(2t)-\frac{z_0^2}{\omega+1}\sin(4t) \biggr ]
\ea
Therefore
\ba
\frac{d\ }{dt} \la H_1 \ra &=& \frac{\epsilon z_0^2}{2} \sin \left ( (2+\epsilon/2)t \right ) 
\nonumber \\ && \hskip -2 cm
+ \frac{\epsilon^2}{16} \biggl [ \left ( 2z_0^2+1 + \frac{2z_0^2}{\omega} (2z_0^2+3) \right ) \sin(2t)
\nonumber \\ && \hskip -1 cm
+ \frac{2\omega z_0^4}{\omega^2-1} \sin(4t)
+\frac{P_1(z_0,\omega)}{\omega(\omega^2-1)}\sin(2(\omega+1)t)
\nonumber \\ && \hskip 0 cm
-\frac{z_0^2P_2(z_0,\omega)}{\omega(\omega^2-1)}\sin(2(\omega-1)t)
\biggr ]
\ea
where
\ba
P_1(z_0,\omega) &=& (4z_0^2+1)(z_0^2+2)\omega^2-2(2z_0^2+1)\omega
\nonumber \\ && \hskip 2 cm
-(2z_0^2+5)z_0^2 \\
P_2(z_0,\omega) &=& (4z_0^2+1) \omega^2-(2z_0^2+1)
\ea

At this level there is no dissipation since energy is simply exchanged back and forth between
the two SHOs. To obtain dissipation we introduce a bath of incoherent, light SHOs. 

\subsection{Bath of light SHOs}
\label{quantumbath}


As in the classical case (see Eq.~(\ref{dotE1defn})), we now integrate over a spectrum of 
incoherent, light SHOs with spectral function $n(\omega )$.
The rate of energy loss of the heavy SHO will be
\ba
{\dot E}_{1,{\rm quantum}} &\equiv& \frac{d\ }{dt} \int_0^\infty d\omega\, n(\omega ) \la H_1 \ra
\nonumber \\ && \hskip -2 cm
= {\rm oscillatory\ terms}
\nonumber \\ && \hskip -2 cm
 - \frac{\epsilon^2}{16} \int_0^\infty d\omega \, n(\omega)\, 
\frac{z_0^2P_2(z_0,\omega)}{\omega(\omega^2-1)}\sin(2(\omega-1)t)
\label{dotE1q}
\ea
We will ignore the non-dissipative oscillating terms. Since $\omega \in [0,\infty)$, the terms
that are not oscillating are the ones that are inversely proportional to $1-\omega$ and whose
oscillation frequency is also $1-\omega$. This means that we only need keep the last term
in Eq.~(\ref{dotE1q}).
We assume that the integral in the last term is dominated by the region $\omega \approx 1$
and take $t \gg 1$ to get
\ba
{\dot E}_{1,{\rm quantum}} &\approx& - \frac{\epsilon^2}{8}  n(1) z_0^4 \, \int_0^\infty d\omega \,
\frac{\sin(2(\omega-1)t)}{2(\omega-1)}
\nonumber \\ 
&\approx& - \frac{\pi}{16}  \epsilon^2 \, n(1) z_0^4 
= - \frac{\pi}{64}  \epsilon^2 \, n(1) X_0^4 
\label{dotE1quantum}
\ea
where we have used the relation $z_0 = X_0/\sqrt{2}$.

\section{Comparison of classical and quantum systems}
\label{comparison}

Comparison of the quantum result in Eq.~(\ref{dotE1quantum}) with the classical result in 
Eq.~(\ref{dotE1classical}) gives
\ba
{\dot E}_{1,{\rm classical}} 
&=& {\dot E}_{1,{\rm quantum}} A^2 \left ( 1- \frac{A^2}{X_0^2} \right ) \nonumber \\
&=& {\dot E}_{1,{\rm quantum}} \frac{E_2}{(\omega/2)} 
\left ( 1- \frac{E_2}{E_1} \right )
\label{cqcompare1}
\ea
where $E_1$ is the energy of the heavy SHO and $E_2$ is the energy
of the light SHO in the bath that is at the resonant frequency $\omega=\Omega$. 
(By rescalings in Sec.~\ref{model} we had set $\Omega=1$.) Next, to determine 
suitable values of $A^2$, equivalently $E_2$, we consider the dynamics of the heavy SHO.

The expectation value of the position of the heavy SHO is given by
\be
\la X \ra = \frac{1}{\sqrt{2}} \sum_{n,m=0}^\infty ( z c_{n+1,m} c^*_{n,m} + z^* c^*_{n+1,m} c_{n,m} ) |f_n|^2
\label{expecX}
\ee
where we used $\sqrt{n}f_n = z f_{n-1}$. This expression will be evaluated to first order in $\epsilon$ in which
case only $c_{n0}$ (not $c_{n2}$) is relevant. From Eq.~(\ref{bn0master}) we write
\ba
c_{n0} = && e^{-i\omega t/2} \biggl [ e^{-i\epsilon (2n+1)t/8} 
\nonumber \\ &&
- i \frac{\epsilon}{8} 
\left \{ z_0^2 e^{-it} + \frac{n(n-1)}{z_0^2} e^{+it} \right \} \sin (t) \biggr ]
\label{cn0master}
\ea
We use Eq.~(\ref{sumovern})
to do the sum over $n$ in Eq.~(\ref{expecX}) and find
\be
z \sum_{n=0}^\infty c_{n+1,0} c^*_{n,0} |f_n|^2
= z_0 e^{-i(1+\epsilon /4)t} - i \frac{\epsilon}{4} z_0 \sin(t)
\ee
Then, to leading order in $\epsilon$
\be
\la X \ra = X_0 \cos \left [ \left ( 1+\frac{\epsilon}{4} \right ) t \right ]
\label{laXraepsilon1}
\ee
which comes from the first term in the square brackets in Eq.~(\ref{cn0master}). The remaining
terms all cancel.

Comparing Eq.~(\ref{laXraepsilon1}) to (\ref{OmegaclassicalN=1}) we see that the classical
and quantum results for the oscillation frequency agree to ${\cal O}(\epsilon)$ if we take
$A = 1$. This is a natural value because then the classical energy ($E_2=1/2$)
is precisely the energy
of the ground state for the light SHO at the resonant frequency $\omega=1$.
Now, with $A=1$,  Eq.~(\ref{cqcompare1}) gives
\be
{\dot E}_{1,{\rm classical}} = {\dot E}_{1,{\rm quantum}} \left ( 1- \frac{\Omega/2}{E_1} \right )
\label{cqcompare2}
\ee
where we have re-inserted $\Omega$, the frequency of the heavy SHO.
The dissipation rates are identical for coherent states with large occupation number
(given by ${\cal N}_1 = E_1/\Omega$) up to ${\cal O}(\epsilon^2)$. 

\section{Conclusions}
\label{conclusions}

Our final results in Eqs.~(\ref{laXraepsilon1}) and (\ref{cqcompare2}) are quite remarkable. 
They show that the classical and quantum oscillation frequencies and dissipation rates 
both agree provided the classical analysis 
is done with the light SHOs in a classical analog of the quantum ground state and if the coherent 
state has large occupation number. This suggests that quantum vacuum dissipation may be studied
classically by giving each of the bath SHOs their ground state energy.

Another surprising conclusion that we mentioned in the introduction is that backreaction on the
classical SHO will make it more quantum. The reason is that the initial coherent state is the most 
classical state, defined by its minimum uncertainty $\Delta X \Delta P =\hbar/2$, and backreaction
can only increase the uncertainty and make the state more quantum. This is opposite of the usual 
role of interactions that cause quantum states to decohere and become more classical. In a
similar way, the initial state is taken to be a product state but it evolves into a mixed state and
the SHOs becomes more entangled with time.

Our calculations are valid only in leading (second) order in perturbation theory. We plan to study the system 
at higher order in perturbation theory and at strong coupling in the future, where the classical and
quantum analyses may deviate from each other. We also plan to study the rate at which the coherent 
state ``incoheres'' due to backreaction, and the rate at which the heavy and light degrees of freedom 
get entangled.

\acknowledgements

I thank Andrei Belitsky, Jeff Hyde, Ted Jacobson, Hank Lamm, Harsh Mathur, Rashmish Mishra, Arif Mohd
and Raman Sundrum for discussions. TV is supported by the U.S. Department of Energy, 
Office of High Energy Physics, under Award No. DE-SC0013605 at ASU.

\appendix

\section{Quantization of SHO in action-angle variables}
\label{quantumphasevariables}

Consider a quantum SHO
\be
H = \frac{p^2}{2} + \frac{x^2}{2} = a^\dag a + \frac{1}{2}
\ee
where
\be
a = \frac{x+ip}{\sqrt{2}}, \ \  a^\dag = \frac{x-ip}{\sqrt{2}}.
\ee
and
\be
[a,a^\dag]=1
\label{aadag}
\ee
follows from $[x,p]=i$.

Now consider the transformation
\be
a = e^{-i\theta} \sqrt{I}, \ \ a^\dag = \sqrt{I} e^{+i\theta}
\label{aadag}
\ee
where we assume that $\sqrt{I}$ is an Hermitian operator and will shortly discuss
the meaning of this operator.
Then
\be
H= I + \frac{1}{2}
\ee
and also Eq.~(\ref{aadag}) leads to,
\be
[\theta, I]=i,
\ee
which has the representation
\be
I = -i \frac{\partial}{\partial\theta}.
\ee
Therefore the normalized eigenstates with energy $n+1/2$ are
\be
\psi_n(\theta) = \frac{e^{in\theta}}{\sqrt{2\pi}}
\ee
with $n=0,1,2,\ldots$ because the wavefunctions are periodic under $\theta \to \theta+2\pi$. 
Eigenstates with negative integer values of $n$ are not allowed in the physical spectrum 
because of the assumed Hermiticity of $\sqrt{I}$ and the definition of $\sqrt{I}$ below.


To interpret $\sqrt{I}$ we define
\be
\sqrt{I} e^{in\theta} = \sqrt{n} e^{in\theta}
\label{sqrtI}
\ee
and work in the basis $\{ e^{in\theta} \}$, assuming that $\sqrt{I}$ acts linearly. 
For example, if
\be
\psi (\theta) = \sum_{n=0}^\infty c_n e^{in\theta}
\ee
where $c_n$ are expansion coefficients, then
\be
\sqrt{I} \psi(\theta) = \sum_{n=0}^\infty c_n \sqrt{n} \, e^{in\theta}
\ee

To recover the usual SHO wavefunctions in position space, we use bra-ket notation.
We would like to go from $\psi_n(\theta) \equiv \la \theta | n \ra$ to 
$\phi_n (x) \equiv \la x | n \ra$.

We write an eigenstate  $|\theta \ra$ of the $\theta$ operator in terms of the energy
basis states as
\be
|\theta \ra = \sum_n |n\ra \la n | \theta \ra
\ee
Then
\be
\la x | {\hat x} |\theta \ra = x \sum_n \la x|n \ra \la n|\theta \ra
\ee
Next we can write
\ba
x \sum_n \la x|n \ra \la n|\theta \ra = 
&& \hskip -0.1 in \la x | {\hat x} |\theta \ra 
\nonumber \\
&& \hskip -1.2 in 
= \frac{1}{\sqrt{2}}\sum_n \la x | 
\left 
[ \exp({-i{\hat \theta}}) \sqrt{I} + \sqrt{I} \exp({+i{\hat \theta}}) \right ] | n\ra \nonumber \\
&& \hskip 2 cm
\times \la n|\theta \ra
\label{xeigenrelation}
\ea
where the hat on $\theta$ emphasizes that it is an operator.
On the right-hand side, we know that $I | n\ra = n |n\ra$ and we take
\be
\sqrt{I} |n\ra = \sqrt{n} |n\ra.
\ee
We also see
\ba
\exp(\pm i {\hat \theta}) |n\ra &=& \int d\theta' \exp(\pm i {\hat \theta})~  |\theta ' \ra \la \theta ' | n \ra \nonumber \\
&& \hskip -1.5 cm 
= \int d\theta' \exp(\pm i {\theta '}) |\theta ' \ra \frac{1}{\sqrt{2\pi}} \exp( i n {\theta '}) \nonumber  \\
&& \hskip -1.5 cm
= \int d\theta'  |\theta ' \ra \la \theta ' | n \pm 1 \ra  \ = | n\pm 1 \ra
\ea
Therefore
\be
\sqrt{I} \exp({\pm i {\hat \theta}}) |n\ra = \sqrt{n\pm 1}\, |n\pm 1\ra .
\ee
Then using these relations in Eq.~(\ref{xeigenrelation}) and with some algebra we obtain 
\be
\sum_n [ x \sqrt{2}\,  \phi_n - \sqrt{n+1}\, \phi_{n+1} - \sqrt{n}\, \phi_{n-1}] \psi_n^*(\theta) =0.
\ee
This relation should hold for all $\theta$ and so
\be
 x \sqrt{2}\,  \phi_n - \sqrt{n+1}\, \phi_{n+1} - \sqrt{n}\, \phi_{n-1} = 0.
\ee
The Hermite polynomials satisfy the recursion relation,
\be 
H_{n+1}(x) -2xH_n (x) + 2n H_{n-1}(x) =0
\ee
leading to the normalized solutions for the SHO wavefunctions in the $x$-representation
\be
\phi_n(x) = \frac{1}{\pi^{1/4}} \frac{1}{\sqrt{2^n n!}} H_n(x) e^{-x^2/2}.
\ee

\bibstyle{aps}
\bibliography{qbackreaction}

\end{document}